# A Resolution of the Puzzle of the Posi-nega Switch Mechanism in the Globally Coupled Map Lattice

T. Shimada and S. Tsukada

*Department of Physics, Meiji University, Higashi-Mita 1-1-1, Kawasaki, Kanagawa 214-8571, Japan*

**Abstract**

We revisit the globally coupled map lattice (GCML). We use a family of universal 'curves of balance' between two conflicting tendencies in the model – the randomness in each map and the coherence due to an averaging interaction – and we locate the posi-nega switch region in the parameter space of the model. We clarify the mechanism of a basic posi-nega switch in the two-cluster regime, which guarantees no mixing of maps across their mean field in the chaotic transient process. We stress that a special attention must be paid for rounding-off errors; otherwise, there fatally occurs an artificial numerical degeneracy of maps in a cluster due to a highly negative Lyapunov exponent of the clustor attractor.

## 1. Introduction

The globally coupled map lattice (GCML) in the simplest form – the homogeneous GCML – is a system of identical $N$ logistic maps coupled via their mean field. It is a representative of the intelligence activity supposed to come from the synchronization among the neurons in the neural network. The GCML evolution equation can be written down in a one-line equation;

$$x_i(t+1) = (1-\varepsilon)f(x_i(t)) + \frac{\varepsilon}{N}\sum_{j=1}^{N} f(x_j(t)), \quad (1)$$

where $t$ is a discrete time and $f(x) = 1 - ax^2$. Yet, GCML is known to have surprisingly rich clustering phases in the space of the nonlinearity parameter $a$ and the coupling $\varepsilon$. In particular in the two-cluster phase, which occurs in the band region in the $(a,\varepsilon)$ plane

$$\varepsilon = 0.08 - 0.16 \text{ at } a = 1.4 \text{ and}$$
$$\varepsilon = 0.24 - 0.33 \text{ at } a = 2.0, \quad (2)$$

it exhibits an interesting posi-nega switch between two clusters of synchronizing maps. The switch found by Kaneko[1] uses an ability of the population ratio $\theta$ ($\theta \equiv N^>/N$ with $N^>$ the number of maps in the majority cluster) as an effective nonlinearity parameter. Before the switch, the maps divide themselves into two clusters oscillating oppositely in phase. The cluster with a higher (lower) average value of maps at even $t$ is called as the positive (negative) cluster. By transporting maps successively from the minority cluster to the majority one, the attractor undergoes successive periodicity doubling. When the unbalance in populations reaches a certain threshold, the system goes into a grand chaotic transient motion. Soon the system comes back to the periodic opposite phase motion. Maps, which were in the positive cluster before the transient steps, are now either all in the positive cluster or all in the negative cluster. The latter case, the half chance, is the posi-nega switch. It is an intriguing phenomenon in that maps behave as if they keep a 'memory' of their former clusters even though they pass through the chaotic transient steps where many channels could open[1].

Already a decade has passed since the discovery, and an application to a more sophisticated case has been also considered[2]. However, the puzzle of the memory has not been elucidated so far. Recently one of the authors (T.S.) has found that the dynamics of the element logistic map – in particular its periodic windows – produces its foliation in the GCML[3]. In this new light, we address ourselves to the old puz-zle. We show that there is a natural mechanism that guarantees no mixing of maps between the clusters. New observations to the two-cluster regime are also presented.

## 2. Cluster Formation

In order to solve the puzzle it is necessary to analyze the cluster attractor dynamics first. The equation (1) is an iteration of two-step process – the mapping and the interaction. In the first, the nonlinearity generally magnifies the variance among the maps. In the second, maps are pulled to the mean field at a fixed rate $1-\varepsilon$. The larger $a$ implies the stronger defocusing of maps, while the larger $\varepsilon$ the stronger focusing. In the two-cluster band (2), the conflict is solved by the opposite phase periodic motion of maps in two clusters.

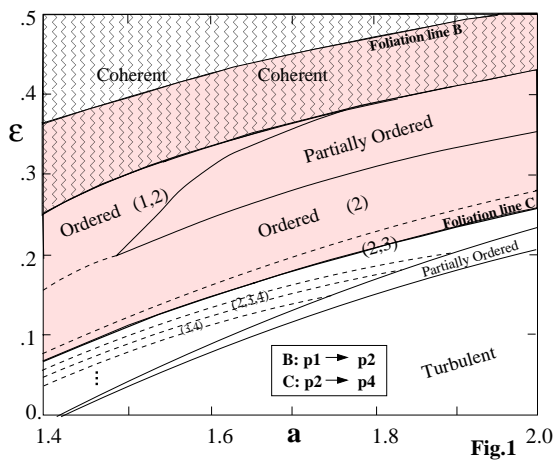

Fig.1

In Fig. 1 we present our prediction (see ref[3]) for the foliation of the period-two region of a single logistic map on the phase diagram. Line B is the foliation of the first bifurcation point, and line C the second. We find that prediction nicely covers the ordered two-cluster regime. It also covers the adjacent part of the coherent phase indicating that the GCML passes through a transient coherent motion with an excess of inputs.

In Fig. 2 we show that the rapid decay of the variations of maps $\delta x_i^\pm \equiv x_i^\pm - x_{av}^\pm$ in each cluster. $N = 50, a = 1.98, \varepsilon = 0.3,$ $\theta = 0.6$. In general, the maps from a random start fall into the two-cluster attractor in the first $\approx 20$ steps and the size of the minority cluster reaches $10^{-40}$ at $\approx 50$ steps.

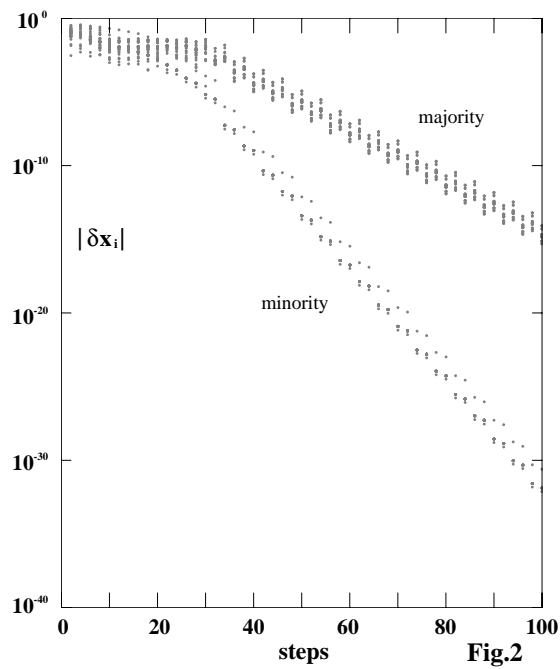

Fig.2

This is a serious warning for the simulation. The maps become unresolved in the usual double precision computation before the system is compelled to the threshold by inputs. With such an artificial degeneracy, the maps will never split again – the memory becomes almost trivially preserved. To avoid such a triviality, we adopt arbitrary precision calculation throughout this note.

Let us analyze the linear stability matrix of GCML following ref[3]. For the cluster attractor configuration there occurs a high degeneracy of the eigenvalues. For the two-cluster attractor in periodicity $p$, there are firstly two eigenvalues,

$$\lambda^{(\pm)} = -2a(1-\varepsilon)\left(\prod_{k=1}^{p} x_{av,k}^\pm\right)^{1/p}, \qquad (3)$$

each with $N^\pm - 1$ – fold degeneracy. These describe the stability of maps in each cluster. There are also two other eigenvalues $\lambda_\pm$ responsible for the stability of the cluster orbits. The observed decay rates of the GCML clusters are plotted in Fig. 3a as a function of $\theta$. These are in an excellent agreement with $\Lambda^{(\pm)} = \log|\lambda^{(\pm)}|$ in Fig. 3b from the matrix-coupled two-map model[1,4] describing the motion of $x_{av}^{\pm}(t)$ – the 'center of mass' of clusters. *The cluster decay rates are solely controlled by the reduced dynamics of the coupled clusters.*

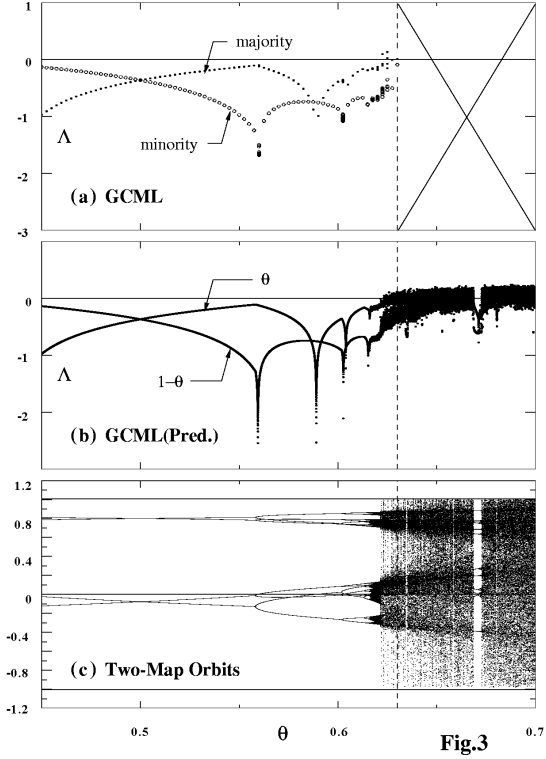

Fig.3

Fig. 3c exhibits the cluster orbits as determined by the matrix model. We find that the orbits successively bifurcate when the cluster decay rates become extreme. On the threshold (the dotted line), the decay almost ceases. This is the start of the grand chaotic motion of the maps. With an excess pulse beyond the threshold, the maps go into the transient steps. The two clusters stop the opposite phase motion and come into a quasi-coherent motion. As there is no stable attractor beyond the threshold, as seen in Fig. 3a, they must soon come back to the original opposite phase motion. Then, the periodic cluster orbits are solely encoded by three parameters $a, \varepsilon$ and $\theta$. Therefore, if any mixing between the clusters has not occurred, *they are destined to follow the same orbits before the transient steps*. If the transient steps are even, all maps should come back to the original clusters; if odd, they should be completely swapped between the positive and the negative cluster – this is the posi-nega switch. (This point is hard to see in the original report[1], where only the orbits at even steps are shown.) Thus the real puzzle boils down to a single question. *What is the mechanism that keeps the population unbalance between two clusters unchanged before and after the chaotic steps?* This question is solved below.

## 3. Transient Process and the No Mixing Mechanism

During the transient time, the clusters move around almost together. Thus, it appears that the maps can easily mix between the clusters. However, there is actually a simple mechanism, which protects them from mixing. The GCML maps evolve under an iteration of a two-step process. From time $t$ to $t+1$, the maps change their value as $x_i \to x'_i \to x''_i$, where the first arrow denotes the mapping, $x'_i \equiv f(x_i)$ and the second the interaction $x''_i \equiv x_i(t+1)$ given by eq. (1). Their mean field changes accordingly as $h \to h' \to h''$, and the variances of maps (the relative coordinates with respect to the mean field) $\delta_i \to \delta'_i \to \delta''_i$. Let us consider the first step. Since all maps are close to the mean field, we obtain $x'_i = f(h + \delta_i) = f(h) + df/dx|_h \delta_i + O(\delta_i^2)$. Averaging this using $\sum_{i=1}^{N} \delta_i = 0$ we find $h' = f(h) + O(\max(\delta_i^2))$. Therefore we find that the new variance is given by

$$\delta'_i \equiv x'_i - h' = df/dx|_h \delta_i + O(\max(\delta_i^2)). \quad (4)$$

The second step is a uniform contraction and the interaction does not affect the mean field, that is, $h'' = h'$, as can be seen by averaging (1).

Hence we see that

$$\delta_i'' = (1-\varepsilon)\delta_i' = J(t)\delta_i \quad (i=1,...,N),$$
$$J(t) = (1-\varepsilon)\,df/dx\,|_h\,. \quad (5)$$

This line is crucial. The factor $J(t)$ is common to all maps. *Thus, at every step in the chaotic transient process, the maps separated by their mean field never mix each other.* We have checked numerically that the maps almost always move in two close clusters around the mean field. But sometimes the clusters split largely. This does not invalidate above resolution. The danger of the mixing occurs when they come close. Our mechanism gives a guarantee of no mixing in such an emergence. Another danger might occur at the cluster configuration $f(x_{av}^+) = f(x_{av}^-)$. But during the chaotic transient steps the clusters evolve almost always together and we have checked that such a coincidence with large splitting is negligible.

We close this note by showing a typical posi-nega switch ($N=50$, a=1.98, $\varepsilon=0.3$) in Fig. 4, which clearly exhibits the memory preserving mechanism by a print circuit pattern. The maps never mix across their mean field during the transient process and the posi-nega switch is a change in the count of the parity (even and odd) of iteration steps.

[1] K. Kaneko, Phys. Rev. Lett. **63**, p. 219 (1989).

[2] K. Kaneko, Physica (Amsterdam) **41D**, p. 137 (1990).

[3] T. Shimada, Technical Report of IEICE, NLP97-159, 71 (1998); T. Shimada and K. Kikuchi, Phys. Rev. **E62**, p. 3497 (2000).

[4] H. Fujigaki, M. Nishi and T. Shimada, Phys. Rev. **E53**, p. 3192 (1996); H. Fujigaki and T. Shimada, *ibid.* **55**, p. 2426 (1997).

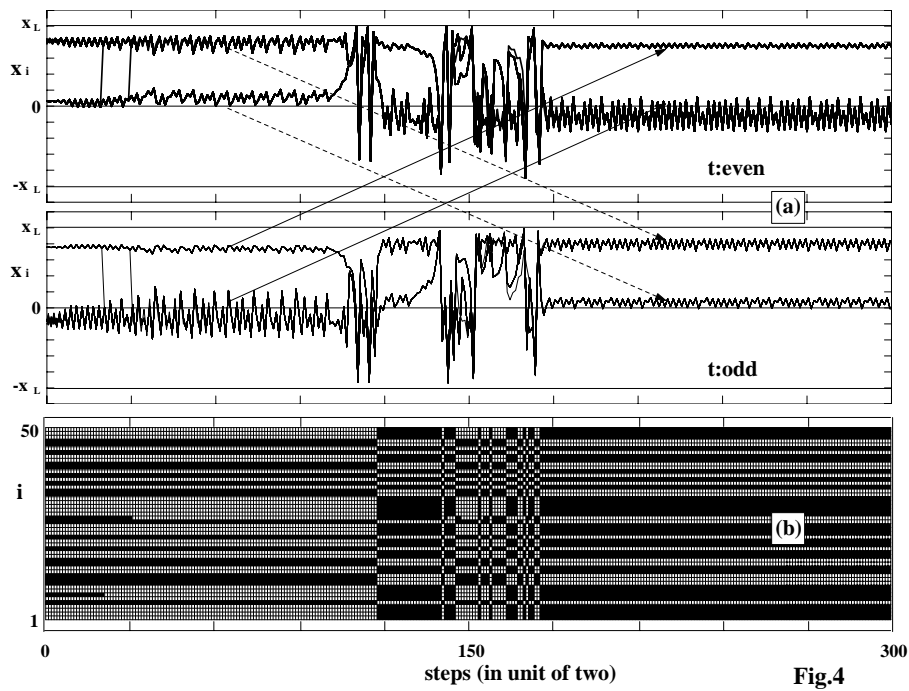

(a) The attractor at even (odd) steps is shown in the upper (lower) diagram. The attractor on the whole, the even and odd iteration steps together, is precisely the same before and after the chaotic transition. The arrows point out that the posi-nega switch is a change in the count of the parity (even and odd) of iteration steps.
(b) The maps are distinguished by their mean field. *The print-circuit shows no mixing of maps across the mean field.*